\documentclass[twocolumn]{aastex631}

\shorttitle{Eruption of the envelope of a massive star}
\shortauthors{Ko et al.}

\graphicspath{{./}{figures/}}

\begin{document}

\title{Eruption of the Envelope of Massive Stars by Energy Injection with Finite Duration}

\author{Takatoshi Ko}
\affiliation{Research Center for the Early Universe (RESCEU), Graduate School of Science, The University of Tokyo, 7-3-1 Hongo, Bunkyo-ku, Tokyo 113-0033, Japan}
\affiliation{Department of Astronomy, Graduate School of Science, The University of Tokyo, Tokyo, Japan}
\author{Daichi Tsuna}
\affiliation{Research Center for the Early Universe (RESCEU), Graduate School of Science, The University of Tokyo, 7-3-1 Hongo, Bunkyo-ku, Tokyo 113-0033, Japan}
\affiliation{Department of Physics, Graduate School of Science, The University of Tokyo, Tokyo, Japan}

\author{Yuki Takei}
\affiliation{Research Center for the Early Universe (RESCEU), Graduate School of Science, The University of Tokyo, 7-3-1 Hongo, Bunkyo-ku, Tokyo 113-0033, Japan}
\affiliation{Department of Astronomy, Graduate School of Science, The University of Tokyo, Tokyo, Japan}
\affiliation{Astrophysical Big Bang Laboratory, RIKEN, 2-1 Hirosawa, Wako, Saitama 351-0198, Japan}
\author{Toshikazu Shigeyama}
\affiliation{Research Center for the Early Universe (RESCEU), Graduate School of Science, The University of Tokyo, 7-3-1 Hongo, Bunkyo-ku, Tokyo 113-0033, Japan}
\affiliation{Department of Astronomy, Graduate School of Science, The University of Tokyo, Tokyo, Japan}
\correspondingauthor{Takatoshi Ko}
\email{ko-takatoshi@resceu.s.u-tokyo.ac.jp}
\begin{abstract}
A significant fraction of supernovae 
show signatures of dense circumstellar material (CSM). While multiple scenarios for creating a dense CSM exist, mass eruption due to injection of energy at the base of the outer envelope is a likely possibility. We carry out radiation hydrodynamical simulations of eruptive mass loss from a typical red supergiant progenitor with initial mass of $15\ M_\odot$, for the first time focusing on the timescale of the injection as well as energy. 
We find that not only sufficient injection energy but also sufficient rate of energy injection per unit time, 
$L_{\rm{min}} \sim 8\times 10^{40}$ erg s$^{-1}$ in this particular model, is required for eruption of unbound CSM. This result suggests that the energy injection rate needs to be greater than the binding energy of the envelope divided by the dynamical timescale for the eruption.
The density profile of the resulting CSM, whose shape was analytically and numerically predicted in the limit of instantaneous energy injection, similarly holds for a finite injection timescale. 
We discuss our findings in the framework of proposed mass outburst scenarios, specifically wave-driven outbursts and common envelope ejection.
\end{abstract}

\keywords{circumstellar matter --- supernovae: general --- stars: mass-loss --- radiative transfer}

\section{Introduction} \label{sec:intro}
Many massive stars explode at the end of stellar evolution, which are observed as supernovae (SNe). SNe display a large variety of spectra and light curves, mainly reflecting the composition of the progenitors and their environments. Although emission lines of SNe are usually broad, some SNe show very narrow hydrogen lines in the spectra. These kinds of SNe are classified as Type IIn, first named by \citet{Schlegel1990}. The existence of the narrow lines strongly suggests that there is a dense circumstellar medium (CSM) slowly moving outside the star.

Assuming that the CSM originates from the SN progenitor, there should be violent mass loss at the late stage of the stellar evolution that forms dense CSM before the SN \citep[e.g.,][]{chugai2004type}. The mass loss rates of SNe IIn progenitors are inferred to be in the range of $0.026-0.12\ \rm{M_{\odot}}\ \rm{yr}^{-1}$ \citep{Kiewe2012}, $10^{-4}-10^{-2}\ \rm{M_{\odot}}\ \rm{yr}^{-1}$ \citep{taddia2013carnegie} or larger than $10^{-3}\ \rm{M_{\odot}}\ \rm{yr}^{-1}$ \citep{moriya2014mass}. These values are much more extreme than standard radiation-driven stellar winds, which generally have mass loss rates of $\dot{M}\lesssim 10^{-5}\ \rm{M_{\odot}}\ \rm{yr}^{-1}$ \citep[e.g.,][]{Vink2001,Smith2014mass}. It is also observationally known that a large fraction of Type IIn SNe progenitors temporarily exhibited outbursts just before ($\sim 1$--$10$ years) the SN explosion \citep[e.g.,][]{Ofek2014}, implying that these events may be responsible for creating the dense CSM.

Despite the increasing number of observations of SNe IIn, the theoretical mechanism of such violent mass loss is not fully understood. There are various scenarios suggested to explain such mass loss. For example, \cite{Smith2014turb} proposed unsteady nuclear burning due to the turbulent convection at the late stage of stellar evolution. \cite{Chevalier2012} proposed the release of gravitational energy due to the common envelope evolution with a binary companion, likely a neutron star or a black hole. \cite{Woosley2007} proposed pulsational pair instability arising from the production of electron–positron pairs in the core of very massive stars. Finally, wave-driven mass loss sourced by nuclear burning in the core is also proposed  {\citep[e.g.,][]{Quataert2012,Shiode2014,Fuller2017,Fuller2018,Leung2020,Wu21,Leung2021_915,Leung2021_923}.}

For many of the models raised here, eruption of the progenitor's envelope is triggered by injection of energy into the bottom of the envelope. Such process was recently studied by \cite{KS20} with radiation hydrodynamical simulations, and \cite{Linial21} with analytical modeling. They showed that there is a minimal injection energy required for mass loss, that depends on the binding energy of the progenitor. However, these calculations have assumed the energy injection to occur instantaneously, i.e. in a much shorter time compared to the envelope's dynamical timescale. Since such an ideal approximation would not always hold, it would be important to consider the effect of having a finite injection duration. A longer duration would make the gravitational pull from the center more important, which would reduce the mass and energy of the eruption. Hence we expect there to be a constraint on the duration on the injection, as well as the energy, for mass eruption to occur.

 {Modelings of continuous energy injection onto the stellar envelope were recently considered by several works \citep[][]{Leung2020,Leung2021_915,Leung2021_923}, in the context of wave-driven mass loss. In \citet{Leung2020}, the amount and asymmetry of the mass loss was investigated by two-dimensional hydrodynamic simulations of hydrogen-rich progenitors. However, the simulations used in their work did not take into account radiation transfer. This may overestimate the mass loss, because loss of energy due to the escape of radiation around shock breakout is not included. In \citet{Leung2021_915}, the dependence of mass loss on the injection energy and duration was investigated taking into account the radiation transfer effect. This work focused on parameterizing the energy injection onto a hydrogen-poor progenitor in order to reproduce a specific Type Ic SN 2018gep. \citet{Leung2021_923} improved the modelling of wave heating and investigated the mass loss, but this work also focused on hydrogen-poor progenitors which are irrelevant to SNe IIn. Moreover, common to these works is that the specific conditions necessary for mass loss to occur were not calculated. %The dependence of mass-loss on the injection energy and the duration in a hydrogen-rich RSG progenitor was systematically investigated to determine the specific conditions necessary for mass eruption to occur.
A model-agnostic criterion of mass loss, as well as its dependence on the injection energy and duration, would be useful to constrain these parameters from observations of SNe with massive CSM and test the proposed models of violent mass loss.
}

In this work, we calculate the mass eruption process with radiation hydrodynamical simulations, varying the duration of the energy injection. We simulate the mass eruption of a red supergiant (RSG) progenitor, and compare the amount of mass loss as well as the final density profile of the resulting CSM. We find that not only a sufficient injection energy but also a sufficiently short timescale, within the order of weeks to months for RSGs, is required.

This paper is organized as follows. In Sect. \ref{sec:method}, we introduce our progenitor model generated with MESA and the simulations done in this work. In Sect. \ref{sec:result}, we present the results of our calculations, with focus on the amount of mass lost. In Sect. \ref{sec:discussion}, we conclude and discuss our results in the context of several proposed mechanisms for mass eruption.

\section{Methods}\label{sec:method}
We use the mass eruption part \citep{KS20} of the open-source code CHIPS \citep{Takei21} to simulate the eruption of the envelope of a massive star years before core-collapse.  In this section we describe the basic setup and key features of our calculations.

\subsection{Progenitor Model}
\begin{deluxetable*}{ccccccc}
\tablenum{1}
\tablecaption{Properties of the progenitor model adopted in this work. The columns are: initial mass, initial metallicity, photospheric radius, effective temperature, mass within helium core, mass of hydrogen envelope, and  {the initial  total binding energy of the outer envelope}.\label{tab:parametar}}
\tablewidth{0pt}
\tablehead{
\colhead{$M_\mathrm{ZAMS}$} & \colhead{ $Z$ }  & \colhead{ $R_\star$ } & \colhead{$T_\mathrm{eff}$} & \colhead{$M_\mathrm{He\ core}$} & \colhead{$M_\mathrm{H\ env}$} & $E_\mathrm{outer}$ \\
\colhead{$[\mathrm{M}_\odot]$} & \colhead{$[\mathrm{Z}_\odot]$} & \colhead{$[\mathrm{R}_\odot]$} & \colhead{[K]}  & \colhead{$[\mathrm{M}_\odot]$}  & \colhead{$[\mathrm{M}_\odot]$} & \colhead{[erg]} 
}
\startdata
15 & 1 & 670 & 4000 & 4.9 & 7.9 & $4.75 \times 10^{47}$
\enddata

\end{deluxetable*}
We adopt a RSG progenitor with an initial mass of 15 $M_\odot$, generated in \cite{Tsuna21} (their R15) using the revision 12778 (\verb|example_make_pre_ccsn| test suite) of MESA \citep{Paxton11,Paxton13,Paxton15,Paxton18,Paxton19} and released as one of the sample progenitor models in CHIPS. The basic parameters of our RSG model are listed in Table \ref{tab:parametar}. As in \cite{KS20}, since we are interested in the (partial) ejection of the envelope, we extract the hydrogen-rich envelope of this progenitor as the computational region. We inject energy into the inner boundary at $r_{\rm inner} \approx 1.7\times 10^{12}$ cm, which correspond to the bottom of the envelope. We use the progenitor that was evolved until core-collapse, but the change in the envelope over the last decade is found to be negligible.

\subsection{1D Radiation Hydrodynamical Simulation}
\begin{deluxetable*}{cc}
\tablenum{2}
\tablecaption{Sets of normalized injected energy $f_{\rm inj}$ and duration of injection $\Delta t_{\rm{inj}}$ simulated in our study.  {There are $31$ parameter sets in total.}\label{tab:initial}}
\tablewidth{0pt}
\tablehead{
\colhead{Injected energy $f_{\rm{inj}}$}&\colhead{Duration of injection $\Delta t_{\rm{inj}}$ [s]}
}
\startdata
0.15,0.16,0.17,0.18,0.19 & $10^3$\\
 {0.2} &  {$10^3,10^5,9\times10^5,10^6$}\\
0.3 &  {$10^4,10^5,10^6,1.8\times10^6,2\times10^6$}\\
0.5 &  {$10^4,10^5,10^6,2\times 10^6, 2.7\times 10^6, 2.9\times 10^6,3\times 10^6,10^{6.5},10^7$} \\
0.8 &  {$10^4,10^5,10^6,2\times 10^6,4\times 10^6,4.8\times10^6,5\times10^6,6\times10^6,7\times10^6,8\times10^6,9\times10^6,10^7$}
\enddata
\end{deluxetable*}
We investigate the violent mass loss that occurs after energy injection, using a radiation hydrodynamics code assuming spherical symmetry \citep{KS20}. We solve the following equations in Lagrangian coordinates:
\begin{equation}
\frac{\partial (1/\rho)}{\partial t} -\frac{\partial (4\pi r^2v)}{\partial m}  = 0,
\end{equation}
\begin{equation}
\frac{\partial v}{\partial t} + 4\pi r^2\frac{\partial p}{\partial m}  = g,
\end{equation}
\begin{equation}
\frac{\partial E}{\partial t} + \frac{\partial (4\pi r^2vp)}{\partial m}  = vg - \frac{\partial L}{\partial m},\label{eqn:energy}
\end{equation}
where $\rho$ is the mass density, $t$ the time from energy injection, $r$ the radius, $v$  the radial velocity, $m$  the enclosed mass, $p$  the pressure,  {$L$ the luminosity, and $E$ the total energy density of radiation, kinetic and internal energy.} The gravitational acceleration $g$ is expressed as
\begin{equation}
g = -\frac{Gm}{r^2}\ ,
\end{equation}
where $G$ is the gravitational constant.

These equations are solved using the piecewise parabolic method \citep{PPM}. In order to calculate $L$ in each cell, we use the diffusion approximation with a flux limiter $\lambda$ \citep{Levermore81}:
\begin{eqnarray}
L &=& -\frac{16\pi^2 a c r^4}{3\kappa}\frac{\partial T^4}{\partial m}\lambda, \\
\lambda&=&\frac{3(2+R)}{6+3R+R^2}, \\
R &=& \frac{\left|\frac{\partial aT^4}{\partial r}\right|}{\kappa\rho a T^4},
\end{eqnarray}
where $a$ is the radiation constant, $c$ the speed of light, $T$  the temperature, and $\kappa$ is the opacity [${\rm cm^2\ g^{-1}}$] calculated in the same way as \citet{KS20} as follows (with dimensional parameters in CGS units):
\begin{equation}
\kappa = \kappa_{\mathrm{molecular}}+ \frac{1}{\frac{1}{\kappa_{\mathrm{H}^{-1}}}+\frac{1}{\kappa_\mathrm{e}+\kappa_{\mathrm{Kramers}}}}, \label{eq:kappa_tot}\label{eqn:kappa_start}
\end{equation}
\begin{equation}
\kappa_{\mathrm{molecular}} = 0.1Z,
\end{equation}
\begin{equation}
\kappa_{\mathrm{H}^{-1}} = 1.1\times 10^{-25}Z^{0.5}\rho^{0.5}T^{7.7},
\end{equation}
\begin{equation}
\kappa_\mathrm{e} = \frac{0.2\left(1+X\right)}{\left(1+2.7\times10^{11}\frac{\rho}{T}\right)\left[1+\left(\frac{T}{4.5\times10^8}\right)^{0.86}\right]},
\end{equation}
\begin{equation}
\kappa_{\mathrm{Kramers}} = 4\times 10^{25}\left(1+X\right)\left(Z+0.001\right)\frac{\rho}{T^{3.5}},\label{eqn:kappa_end}
\end{equation}
where $\kappa_{\mathrm{molecular}}$ denotes the molecular opacity, $\kappa_{\mathrm{H}^{-1}}$ the hydrogen opacity, $\kappa_\mathrm{e}$ the electron scattering opacity, and $\kappa_{\mathrm{Kramers}}$ Kramers' opacity for absorption regarding free-free, bound-free, and bound-bound transitions. Parameters $X$ and $Z$ are the mass fractions of hydrogen and elements heavier than helium. We integrate the energy conservation equation (\ref{eqn:energy}) with respect to time in a partially implicit manner \citep{Shigeyama90}. 

The HELMHOLTZ equation of state\footnote{ {The electron-positron pair contribution is included in the HELMHOLTZ equation of state. However this work focuses on the outer hydrogen envelope of the star, where the electron-positron pair contribution is negligible due to the low temperature ($\lesssim 10^{5}$ K).}} \citep{Helmholtz2000} is used to calculate thermodynamic quantities from  {the internal energy density and $\rho$}, which are obtained from the time integration of the hydrodynamical equations. For the range not covered by this equation of state, the pressure is assumed to be the sum of that of ideal gas and that of black-body radiation, i.e.,
\begin{equation}
p = \frac{\mathcal{R}}{\mu}\rho T + \frac{a}{3} T^4,
\end{equation}
where $\mathcal{R}$ denotes the gas constant, and $\mu$ the mean molecular weight.

When solving the hydrodynamical equations, the boundary conditions are given as follows:
\begin{equation}
v_{\mathrm{inner}} = 0,\ r_{\mathrm{inner}} = 1.657\times10^{12}\ \mathrm{cm},\ p_{\mathrm{outer}}=0.
\end{equation}
Here the subscript inner (outer) indicates quantities at the base (outer edge) of the envelope. In contrast to \cite{KS20} that considered a fixed energy injection timescale much shorter than the envelope's dynamical time, we consider the injection timescale as a free parameter. The energy is injected at a constant rate as
\begin{equation}
\frac{\mathrm{d} E}{\mathrm{d} t} = \frac{E_{\rm inj}}{\Delta t_{\rm{inj}}} \equiv \frac{f_{\rm inj}E_{\rm outer}}{\Delta t_{\rm{inj}}},
\end{equation}
where $E_{\rm inj}$ is the total injected energy.  {The energy is equally injected in the innermost $10$ cells of the computational region.
} Following \cite{Tsuna21} and \cite{Takei21}, we characterize $E_{\rm inj}$ by the parameter $f_{\rm inj}$, which is the ratio of $E_{\rm inj}$ to the  {initial} total binding energy of the outer envelope $E_{\rm outer}$ in Table \ref{tab:parametar}. The earlier works by \cite{KS20} found that for RSGs $f_{\rm inj}$ of a few $10\%$ or more results in (partial) ejection of the envelope. We investigate $22$ different parameter sets of $(f_{\rm inj}, \Delta t_{\rm{inj}})$ in this range, as shown in Table \ref{tab:initial}, and follow the hydrodynamical evolution of the envelope for $5$ years.
 
\section{Results}\label{sec:result}
\begin{figure*}
 \centering
 \includegraphics[width=1.1\linewidth]{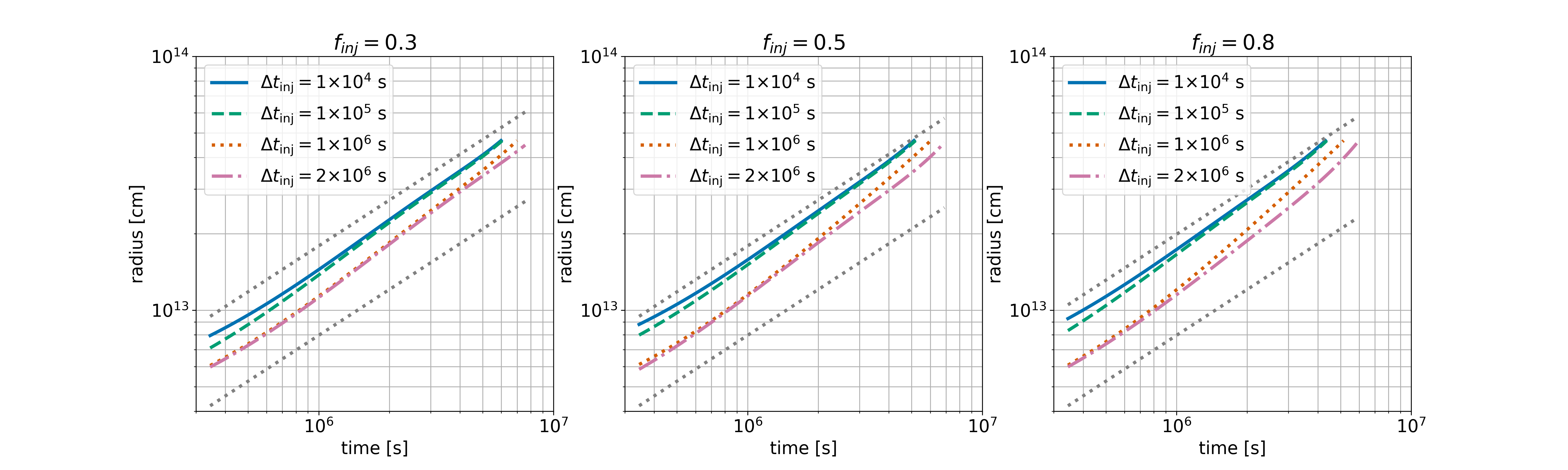}
\caption{Time evolution of the shock propagating inside the envelope until breakout. The curves are obtained by fitting sparsely distributed data points extracted from simulations with a third-order polynomial.  { Overlaid as dotted lines is the relation $r_{\rm{shock}}\propto t^{0.6}$, predicted from the Sedov-Taylor solution for a blast wave propagating through a medium of density profile $\rho\propto r^{-1.7}$.}}
 \label{fig:shocl_evo}
 \end{figure*}
 
The energy injection generates an outward shock wave that propagates in the envelope of the progenitor \citep[see e.g. Fig. 3 of][]{KS20}. When the shock approaches the star's surface it breaks out, and a part of the envelope is ejected to form a CSM. In this section we present  results of our simulations, focusing on the shock propagation, mass ejection and the final density profile of the CSM.

\subsection{Shock Propagation in the Star}\label{sec:shock_propa}

We extract the time evolution of the shock from the output files of the CHIPS code\footnote{\texttt{resultXX.txt}, where XX is an integer from 00 to 99.} that record the hydrodynamical quantities at each cell. The evolution of the shocks for various parameter sets is shown in Figure \ref{fig:shocl_evo}. We have fit the radius of the shock as a function of time $r_{\rm shock}(t)$ with a third-order polynomial using the \verb|numpy.polyfit| module, which results in an error within 0.3\% in radius. 

Due to the energy injection a shock is formed and accelerated near the base of the envelope, but it gradually decelerates as more material is swept up as shown in Figure \ref{fig:shocl_evo}.  {Figure \ref{fig:shocl_evo} shows that the shock propagation in the star initially follows a power law of  $r_{\rm shock} \propto t^{0.6}$, especially for $\Delta t_\mathrm{inj}$'s much shorter than the dynamical timescale of the envelope
\begin{eqnarray}\label{eq:t_dyn}
    t_{\rm dyn} &\approx& \sqrt{\frac{R_*^3}{GM_*}} \sim 8\times 10^6\ {\rm s}\left(\frac{R_*}{670R_\odot}\right)^{3/2}\left(\frac{M_*}{13M_\odot}\right)^{-1/2}.
\end{eqnarray}
%which is found to be reproduced by the Sedov-Taylor self-similar solution \citep[][]{Taylor1950,Sedov1959}. 
This dependence is found to be consistent with the Sedov-Taylor self-similar solution \citep[][]{Taylor1950,Sedov1959}, where the expansion of the blast wave for our initial RSG model (with a density profile of approximately $\rho \propto r^{-1.7}$ near the inner boundary) should follow $r_{\rm shock} \propto t^{0.61}$.}

As the shock approaches the surface, the upstream density quickly decreases and the shock is slightly re-accelerated. Eventually the shock reaches the breakout radius, where the optical depth $\tau$ satisfies $v_{\rm shock}=dr_{\rm shock}/dt\sim c/3\tau$ and the internal energy stored in the shock downstream starts to be lost by radiative diffusion. This happens 1--3 months after energy injection, depending on the model parameters.

% {When there is a shock cooling at the stellar envelope, it should lead to a distinguishable luminosity peak. In this work, the luminosity of this brightening is almost $10^{39}$ erg s$^{-1}$, which is fainter than what is observed. However, as written in \cite{KS20}, a luminosity of $10^{39}$ erg s$^{-1}$ cannot be observed unless it occurs within $\lesssim 10$ Mpc, and such a brightening within that range is very rare, so the results of this study are not inconsistent with the observations.}
\subsection{Mass Eruption}\label{sec:mass-loss}

 \begin{figure}
 \centering
 \includegraphics[width=\linewidth]{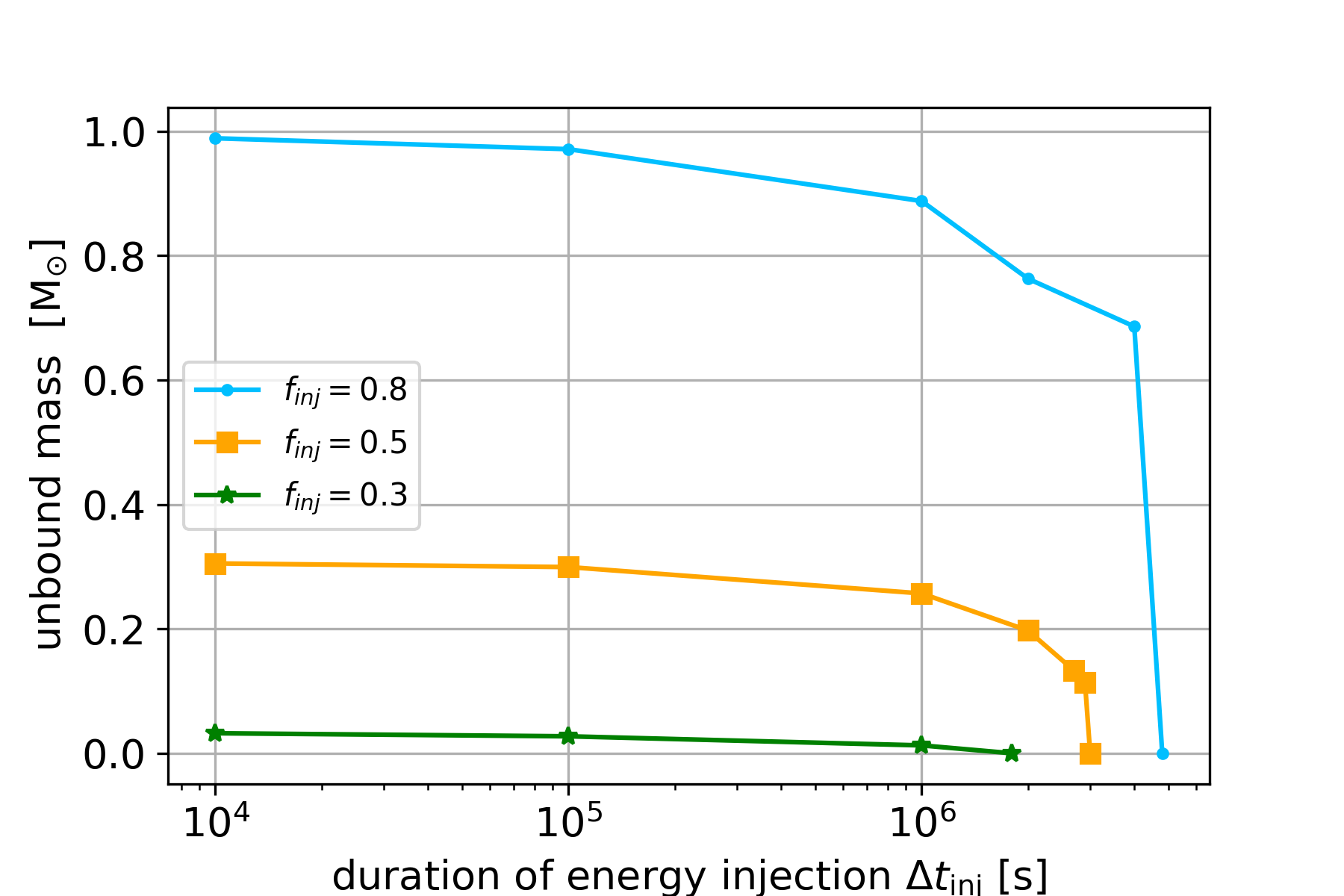}
\caption{The mass of unbound matter for different values of  $f_{\rm{ inj}}$ and $\Delta t_{\rm{inj}}$ calculated by the radiation hydrodynamical simulation.}
 \label{fig:mass_erupt}
 \end{figure}
  \begin{figure}
 \centering
 \includegraphics[width=\linewidth]{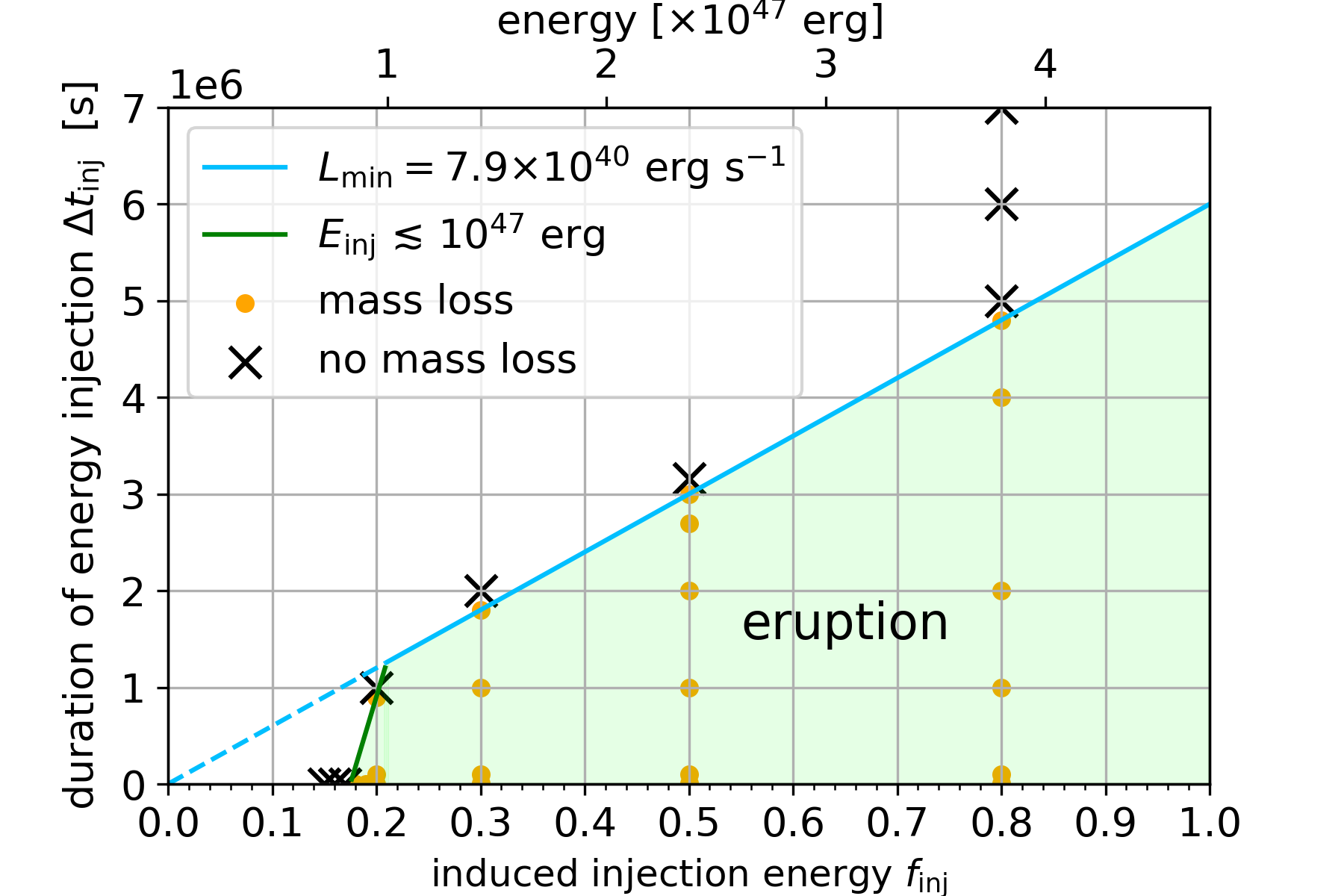}
\caption{Relation between $f_{\rm{inj}}$ and $\Delta t_{\rm{inj,max}}$ obtained from our simulations (orange dots  {and black crosses}).  {The orange points (black crosses) are where unbound CSM is found to form (not form) in our calculations. The relation is approximately linear (blue line) at large $f_{\rm inj}$ with slope $\Delta t_{\rm{inj,max}}/f_{\rm{inj}}=6\times 10^6$ [s], while it transitions at $f_{\rm inj}\approx 0.2$ ($E_{\rm inj}\approx 10^{47}$ erg), the minimum $f_{\rm inj}$ required for mass loss.} We predict mass eruption to occur only in the shaded region of the parameter space.}
 \label{fig:tau_max}
 \end{figure}
 
The material released from the star upon shock breakout expands and reduces the pressure by converting its internal energy to kinetic energy. When the pressure in the material becomes negligible, the motion is exclusively controlled  by the central star's gravity. At this point, the material is composed of an unbound part  at the outer region and a bound part at the inner region that eventually falls back towards the central star. Thus we focus on the unbound CSM.

Figure \ref{fig:mass_erupt} shows the dependence of the mass of the unbound material on $\Delta t_{\rm inj}$, obtained from our simulations in the range $f_{\rm{inj}}=0.3$--$0.8$. It can be seen that while the mass is rather insensitive to $\Delta t_{\rm{inj}}$ at low values, the energy injection with a $\Delta t_{\rm{inj}}$ longer than a few $\times 10^6$ s significantly reduces the amount of ejected envelope. This is because the gravitational force from the center becomes important when $\Delta t_{\rm inj}$ is comparable to the dynamical timescale at the outer edge of the hydrogen envelope (see Eq. (\ref{eq:t_dyn})).
A longer $\Delta t_{\rm inj}$ would result in deceleration of the shock, hence a lower amount of mass to be accelerated to velocities sufficient for being gravitationally unbound.

Since the total energy of the envelope after energy injection $(1-f_{\rm inj})(-E_{\rm outer})$ is still negative, a very long $\Delta t_{\rm inj}$ should result in no mass ejection. As shown in Figure \ref{fig:mass_erupt}, we find that the maximum value $\Delta t_{\rm{inj,max}}$ of $\Delta t_{\rm inj}$ for mass eruption, is of order months and shorter for smaller $f_{\rm inj}$. Figure \ref{fig:tau_max} shows the relation between $f_{\rm{inj}}$ and $\Delta t_{\rm{inj,max}}$, and illustrates that there is a minimum luminosity required for mass loss to occur. The relation is linear with a slope of $\Delta t_{\rm{inj,max}}/f_{\rm{inj}} = 6\times10^6$ s, which is close to the dynamical time scale $t_{\rm{dyn}}\sim 8\times 10^6$ s. This suggests that the following equation roughly holds:
\begin{equation}
    \frac{E_{\rm{inj}}}{\Delta t_{\rm{inj,max}}} \sim \frac{E_{\rm{outer}}}{t_{\rm{dyn}}}.
    \label{eq:Linj}
\end{equation}
In the case of our RSG progenitor, the constraint of $\Delta t_{\rm{inj,max}}/f_{\rm{inj}}$ for mass loss corresponds to an injection luminosity of $7.9\times10^{40}$ erg s$^{-1}$.

 {In order for the shock to reach the surface of the star, a certain amount of energy is required, because some of the energy involved in the shock is lost due to ascending the gravitational potential during its propagation and escape of radiation around breakout. This condition should appear as a different component from the blue line in Figure \ref{fig:tau_max}. \cite{Linial21} and \cite{KS20} found that there exists a minimal energy required for RSG progenitors to achieve eruptive mass loss, of the order of $10^{46}$--$10^{47}$ erg. We repeat the simulation for lower values of $f_{\rm inj}$, with an instantaneous energy injection of $\Delta t_{\rm inj}=10^3$ s and check the existence of minimum injection energy ($E_{\rm inj}\lesssim 8\times 10^{46}\ {\rm erg}$) as shown in Figure \ref{fig:tau_max} which is roughly consistent with the both works. By simulating $f_{\rm inj}$ just above this lower limit, we additionally found that this appears as a hard cutoff near the lower limit, shown by the green line in Figure \ref{fig:tau_max}.}

We conclude that there exists both a minimum energy and minimum luminosity for mass loss to occur. In Sect. \ref{sec:discussion} we consider the implications of our results in the context of several models proposed for mass eruption.

\begin{figure}
 \centering
 \includegraphics[width=\linewidth]{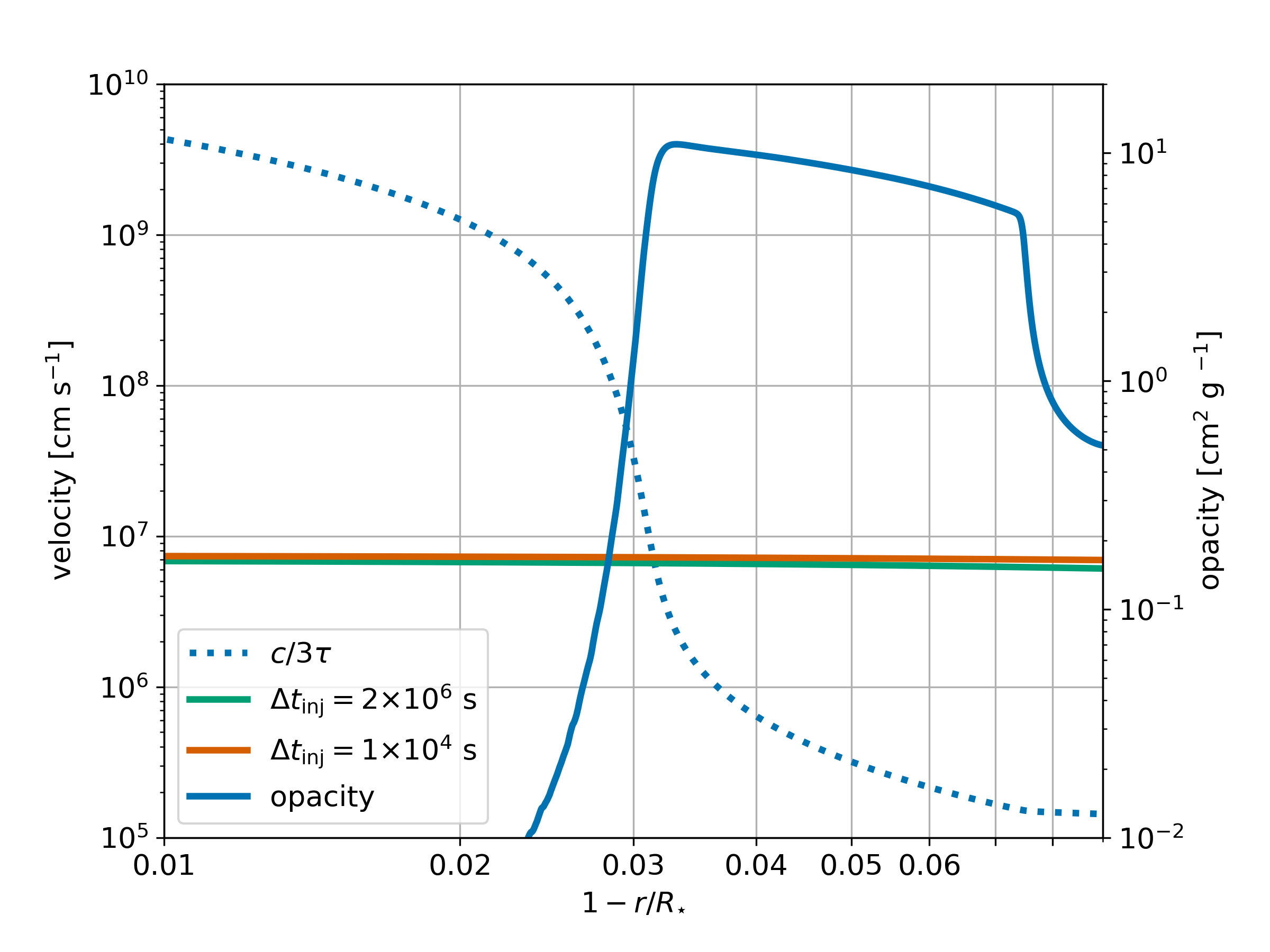}
\caption{The shock and photon diffusion velocities ($c/3\tau$) and opacity near the photosphere of the star as functions of the normalized depth from the pohotosphere. The two shocks are generated with the same $f_\mathrm{inj}=0.5$ but with different $\Delta t_\mathrm{inj}=10^4$ s and $2\times10^6$ s. The opacity is calculated from equation (\ref{eq:kappa_tot}) using the density and temperature profile of the star just before shock breakout.}
 \label{fig:breakout}
 \end{figure}

\subsection{Analytical Investigations of Erupted Mass}\label{sec:semi-ana}
The value of the mass of the erupted material can not only be obtained by numerically following the mass eruption for many years like done above, but can also be roughly inferred by simpler analytical approaches. We introduce such inferences and check the consistency between the numerical results.

First, there exists a lower limit of the ejecta mass, depending only on the progenitor, because mass ejection occurs only when the material at and beyond the breakout radius swept up by the shock can obtain a velocity exceeding its escape velocity \citep{Linial21}. 
Assuming that the opacity $\kappa$ is constant near the surface, this mass is expressed as
\begin{equation}
    m_{\mathrm{min}} \approx \frac{4\pi C}{3\sqrt{2}}\left(\frac{c^2R_*^5}{G\kappa^2M_*}\right)^{1/2},\label{eqn:mineq}
\end{equation}
where $C\approx 2$. This coefficient $C$ is actually a function of the depth from the photosphere (see also Eq. (\ref{eqn:const})). 

For the outer layer of RSGs, the opacity is however sensitive to the temperature due to the partial ionization of hydrogen. Figure \ref{fig:breakout} shows the radial dependence of  {the shock velocity, photon diffusion velocity and the opacity $\kappa$ (Eq. (\ref{eqn:kappa_start}))}, around the breakout radius where $\tau\sim c/3v_{\rm shock}$.

The shock velocity (nearly horizontal solid lines) is of the order of $100\ {\rm km\ s^{-1}}$ for all cases, and thus the breakout radius is close to the photosphere ($r \approx 0.97 R_\star$). Around this radius the opacity rapidly changes from $\approx 10\ {\rm cm^2\ g^{-1}}$ to  $< 1\ {\rm cm^2\ g^{-1}}$. If we set $\kappa=1\ {\rm cm^2\ g^{-1}}$ as a fiducial value, we then obtain the lower limit of the mass as 
\begin{equation}
    m_{\mathrm{min}} \sim 3 \times 10^{-2} \ \mathrm{M}_{\odot}\left(\frac{C}{2}\right)\left(\frac{R_*}{670R_\odot}\right)^{5/2}\left(\frac{M_*}{13M_\odot}\right)^{-1/2}.
\end{equation}
This result is consistent with the mass loss from the hydrodynamics simulations, where we obtain the lowest finite mass of the unbound CSM to be $3.2\times 10^{-2} M_\odot$ from a model with a parameter set of $f_{\rm inj}=0.3$ and $\Delta t_{\rm inj}= 1\times10^6$ s.

Second, one can infer the unbound mass from the evolution of the shock velocity depicted in Figure \ref{fig:shocl_evo}. The material that crosses the shock is heated, accelerated, and acquires an outward velocity upon mass eruption. The final speed $v_\mathrm{eje, fin}$ of this material  is determined by the structure of the envelope as well as the shock velocity, and can be written as \citep[e.g.,][]{Matzner99,Ro13}
\begin{equation}
    v_\mathrm{eje,fin} = C(x_0)\times v_{\rm shock}(x_0),
    \label{eq:v_fin}
\end{equation}
where both of $C$ and $v_{\rm shock}$ are functions of the normalized depth from the photosphere of the star $x_0=1-r_{\rm shock}/R_*$. For an envelope with a polytropic index $3/2$ appropriate for RSGs, an approximate fit to $C(x_0)$ is given as
\begin{equation}
    C(x_0) \approx 2.1649 \left(1-0.51x_0^{1/3}+0.76x_0^{2/3}-1.19x_0\right),\label{eqn:const}
\end{equation}
which we note is independent of our model parameters. From our polynomial fit of $r_{\rm shock}(t)$, we can obtain $v_{\rm shock}=dr_{\rm shock}/dt$ and thus $v_{\rm eje, fin}(x_0)$. An important assumption is that gravity from the central star is neglected when determining the parameter $C$. We can nevertheless compare this $v_{\rm eje, fin}(x_0)$ with the escape velocity $v_{\rm esc}=\sqrt{2Gm/r}$ at the same depth, and determine whether (and roughly how much) mass can be ejected.

\begin{figure}
 \centering
 \includegraphics[width=\linewidth]{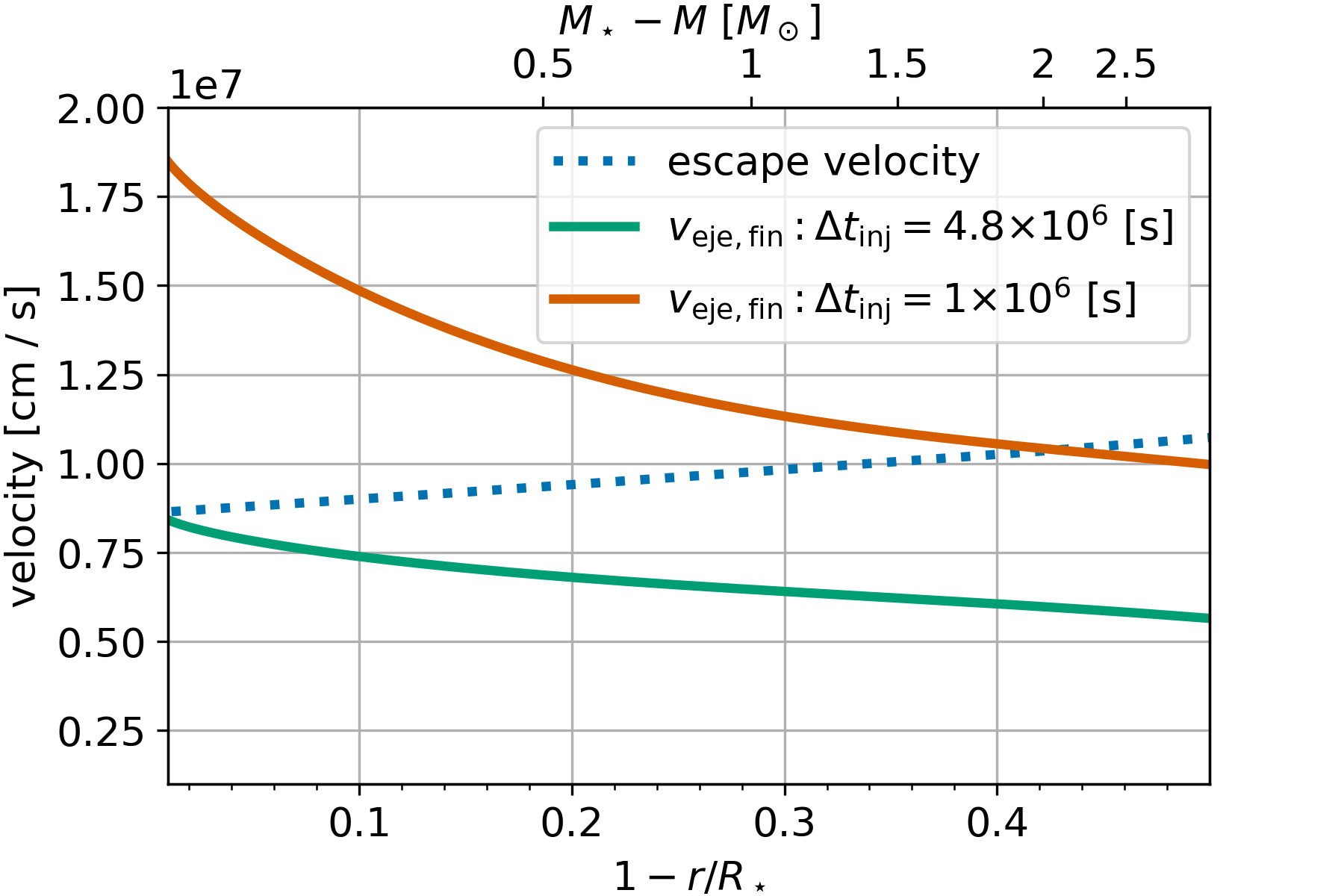}
\caption{The final velocities of the ejecta (solid line) and the escape velocity (dashed line) as functions of the normalized depth from the photosphere. The former velocities are calculated using equation (\ref{eq:v_fin}), for two parameter sets with $f_{\rm{inj}}=0.8$.
}
\label{fig:vel}
 \end{figure}

To illustrate this, Figure \ref{fig:vel} shows a comparison between $v_{\rm eje,fin}$ (solid lines) and $v_{\rm esc}$ (dotted line) for two parameter sets. The figure indicates that a longer $\Delta t_{\rm{inj}}$ would result a narrower region where $v_\mathrm{eje,fin}$ exceeds $v_{\rm{esc}}$, i.e. a more gentle mass loss as seen in Sect. \ref{sec:mass-loss}. From the figure the lower solid line of $\Delta t_{\rm inj}=4.8\times10^6$ sec results in no unbound CSM, while the upper solid line with shorter $\Delta t_{\rm inj}=10^6$ sec results in about $2M_\odot$ of unbound CSM. These are consistent with the results obtained from numerical simulations within a factor of a few.

However, we note that this method may give inaccurate estimates of the unbound CSM mass when $v_{\rm eje,min}$ is marginally greater than $v_{\rm esc}$ and only a little mass close to the breakout radius is ejected. One reason is that the calculation of $C(x_0)$ assumes adiabatic expansion. This might lead to an overestimation if radiation can efficiently escape from the shock downstream. Another issue may arise from a small but finite error in the polynomial fitting for obtaining $v_{\rm shock}$. In our fitting, $r_{\rm{shock}}$ was found to have an error of a few $\times 10^{10}$ cm, and the interval of the time sample for the fitting is $\approx 3\times 10^5$ s, thus $v_{\rm{shock}}$ contains an error of $\sim 1\ \rm{km}\ \rm{s}^{-1}$ at each timestep. The escape velocity is somewhat insensitive to the mass measured from the surface, with a relative change of $1\ {\rm km\ s^{-1}}$ in $\sim 10^{-2}\ M_\odot$. This fitting error thus prevents this method from accurately obtaining the CSM mass if it is around this resolution.

 {Third, the amount of unbound matter due to the wave-driven mass loss is analytically calculated in \cite{Matzner2021}, neglecting energy loss due to radiation and assuming that the injection time is much shorter than the dynamical timescale. \cite{Matzner2021} found that the amount of the unbound matter is proportional to $f_{\rm{inj}}^{\theta}$. Here, $\theta$ is expressed as
\begin{equation}
    \theta = \frac{4\gamma(1+2\beta n)}{3\beta(5+2n)},
\end{equation}
where $\beta\approx 0.19$, $\gamma$ is the adiabatic index, and $n$ is the polytropic index. Adopting $(\gamma,n)=(5/3,3/2)$ appropriate for RSG models, we obtain $\theta\approx 2.09$. For the case where $\Delta t_{\rm inj}\ll t_{\rm dyn}$, the combinations of injection energy and the amount of the unbound matter in our work $(f_{\rm{inj}},\ {\rm unbound\ mass})$ are $(0.8,0.99M_{\odot})$, $(0.5,0.31M_{\odot})$ and $(0.3,0.032M_{\odot})$. Our results at $f_{\rm inj}=0.5$ $(0.3)$ predict a factor of $1.2$ $(4.0)$ lower mass than that predicted from the dependence $\propto f_{\rm{inj}}^{2.09}$ normalized at $f_{\rm{inj}}=0.8$. We presume this to be because the smaller the injection energy, the more significant is the energy loss due to radiation.
}

\subsection{Final Density  {and Velocity} Profiles of the CSM}

\begin{figure}
 \centering
 \includegraphics[width=\linewidth]{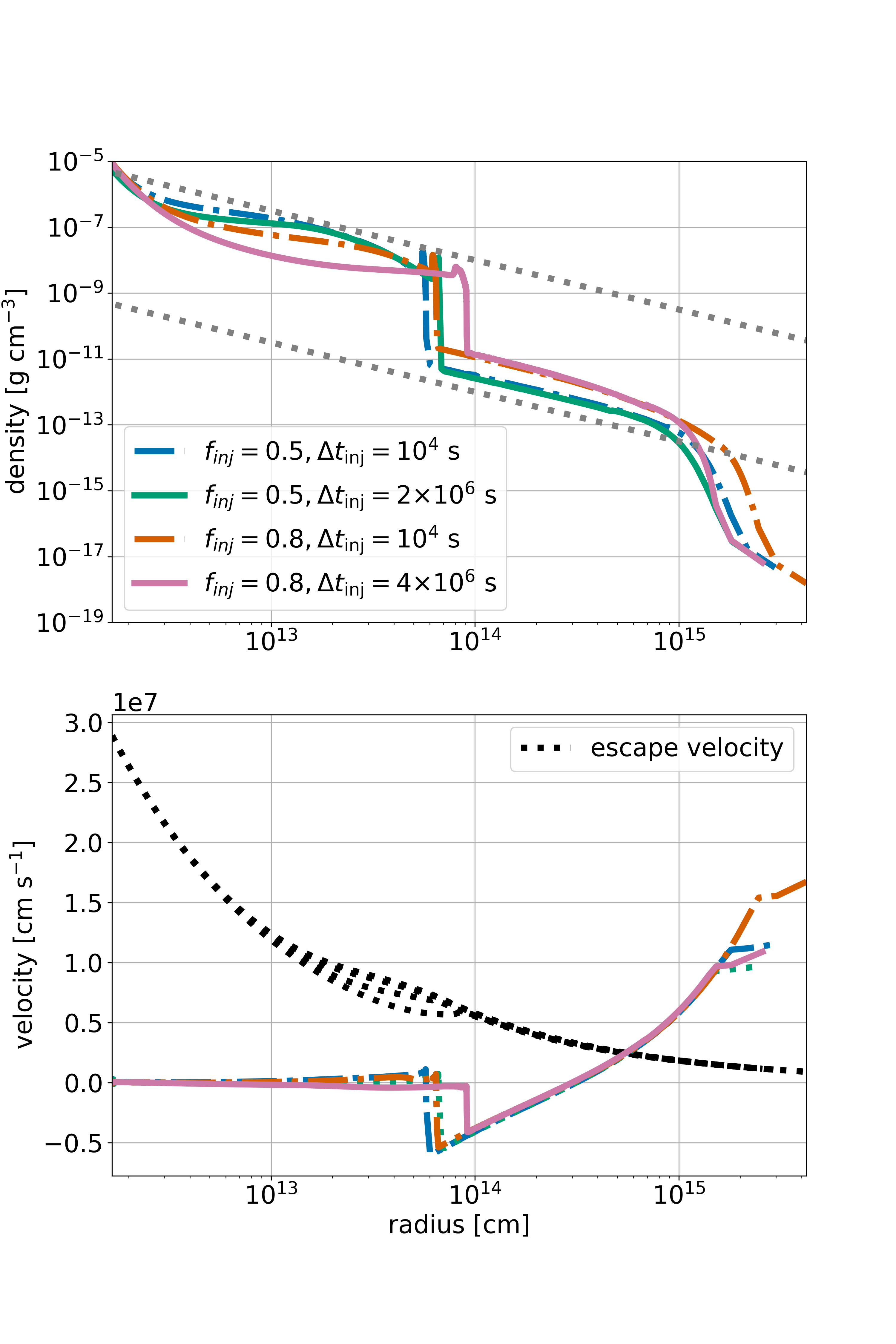}
\caption{(Upper panel) Density profile of the CSM at 5 years after the energy injection, for different parameter sets of $f_{\rm inj}$ and $\Delta t_{\rm{inj}}$. Overlaid as dotted lines are the relation $\rho\propto r^{-1.5}$, predicted in previous studies for the inner part of the dense CSM. (Lower panel) Corresponding velocity profiles for the same parameter sets. The dotted lines show the escape velocity, which is independent of the model parameters in the CSM region ($\gtrsim 10^{14}$ cm).}
 \label{fig:density}
 \end{figure}
 
The density profiles of the final CSM for several parameter sets are shown in the top panel of Figure \ref{fig:density}. When $f_{\rm inj}$ is fixed, the CSM generated with a shorter $\Delta t_{\rm inj}$ extends to larger distances, due to the slightly larger shock speed around shock breakout.

The shape of the density profile appears to be a double power law with the inner region following $\rho \propto r^{-1.5}$ (solid lines in Figure \ref{fig:density}), in agreement with that derived analytically in \cite{Tsuna21} and numerically in \cite{KS20} in the case of instantaneous energy injection. The inner power law originates from the fallback of the bound part of the CSM, which is rather insensitive to the energy injection history. The discontinuity in Figure \ref{fig:density} at $r\sim 10^{14}$ cm is the shock created by the collision of the falling CSM and the progenitor. We thus conclude that this density profile of the CSM is a robust outcome independent of the mass eruption mechanism, given that a sufficient time passes for the bound CSM to start falling back.

 {The velocity profile of the final CSM for several parameter sets are shown in the bottom panel of Figure \ref{fig:density}. The outermost velocities are around $100$ km s$^{-1}$, slightly larger than the escape velocity of the original RSG progenitor. This is around the lower end of the observed CSM velocity from H$\alpha$ lines of SNe IIn, of about $100$--$1000$ km s$^{-1}$ \citep[][]{Kiewe2012,Taddis2013}. We note that the observed CSM velocity may be affected by acceleration due to radiation emitted in the previous phase of the SN. Alternatively, a more compact hydrogen-rich progenitor like blue supergiants may be responsible for a fraction of the explosions with highest CSM velocities.}%This faster velocity is thought to be due to acceleration by radiation from the shocked region. Thus, the wind velocity calculated in this work is consistent with the observation.}

\section{Discussion and Conclusion}
\label{sec:discussion}
In this paper, we investigated how eruptive mass loss of a massive star's envelope is affected with the timescale of the energy injection at its base. We found that not only sufficient energy is required for eruptive mass loss, but also that the energy must be released within a certain time period, as shown in Figure \ref{fig:tau_max}.  {This is qualitatively consistent with \cite{Leung2021_915}, which put limitation on the energy injection for mass-loss to occur for a hydrogen-free progenitor.} This limits the injection luminosity to be $L_{\rm{min}} \gtrsim 8\times 10^{40}$ erg s$^{-1}$ to cause mass loss for our typical RSG stellar model.  For an injection energy of $\sim 10^{47}$ erg, this limits the injection timescale to be within a few weeks to reproduce a IIn-like CSM.

\subsection{Application to Proposed Mass-loss Mechanisms}
We apply our results to several proposed mass loss mechanisms. Recently \cite{Wu21} examined pre-SN mass loss due to heating of the envelope by gravity waves excited at the inner core. Their calculations for progenitor models with initial masses of $\lesssim 15$M$_\odot$ show that during the neon/oxygen burning phase, an energy of $10^{47}$ erg is supplied with a luminosity of $10^{41}$--$10^{42}$ erg s$^{-1}$, i.e. an energy injection timescale of $10^5$--$10^6$ s. Comparing this with Figure \ref{fig:tau_max}, we find that this energy injection can indeed result in eruptive mass loss.  {However, it should be noted that for wave-driven mass loss the energy dissipation occurs near the base of the star’s outer envelope. Uncertainties in the position of the energy deposition not considered in our work can alter the binding energy, and hence can largely affect the conditions for mass eruption.}

Mass ejection may also occur through binary interactions, e.g. when the massive star forms a common envelope with a compact object (CO). Such mechanism has been discussed as the origin of the massive CSM \citep[e.g.,][]{Chevalier2012},  {and may be followed by an explosion, with energy comparable to or even exceeding SNe, when the CO further spirals into the star's helium core \citep[e.g.,][]{Fryer1998,Barkov2011}}. 
% {It is true that in order for the CSM to be generated in the vicinity of the star, the star needs to be already close to its collapse when the companion star ejects the common envelope, where the overlap of time frame is extremely short in the evolution timescale.  However, it has been discussed that high-energy explosions are caused by the common envelope evolution with compact objects \citep[][]{Fryer1998,Barkov2011}, and interaction powered supernovae are expected to occur when such high-energy explosions are triggered.} 
When a CO spirals in the common envelope, the orbital energy between the CO and the star's core can be (partially) used to expel the envelope. The fraction of the dissipated energy that can be used for envelope ejection (the $\alpha$ parameter; \citealt{Livio88,Ivanova13}) is expected to be $\alpha<0.7$, or even $\alpha \sim 0.1$ \citep[][and references therein]{Klencki2021}. Assuming that the CO spirals in at the base of the envelope and the orbital energy $E_{\rm orb}$ is dissipated within roughly the period $t_{\rm Kep}$ of the Kepler motion there, the energy injection rate ${E_{\rm{inj}}}/{\Delta t_{\rm{inj}}}$ can be crudely expressed as
\begin{equation}
     \frac{E_{\rm{inj}}}{\Delta t_{\rm{inj}}}\sim \frac{\alpha E_{\rm{orb}}}{t_{\rm{Kep}}}.
\end{equation}
The orbital quantities scale as
\begin{eqnarray}
E_{\rm orb} &\approx& G\frac{M_{\rm He}M_{\rm CO}}{2r_{\rm inner}} \nonumber \\
&\sim& 7\times 10^{47}\ {\rm erg} \left(\frac{M_{\rm He}}{5\ M_\odot}\right)\left(\frac{M_{\rm CO}}{1\ M_\odot}\right)\left(\frac{r_{\rm inner}}{10^{12}\ {\rm cm}}\right)^{-1},\\
t_{\rm Kep} &\approx& 2\pi \left[\frac{r_{\rm inner}^3}{G(M_{\rm He}+M_{\rm CO})}\right]^{1/2} \nonumber \\ 
&\sim&2\times10^{5}\ {\rm s}\left(\frac{M_{\rm He}+M_{\rm CO}}{6\ M_\odot}\right)^{-1/2}\left(\frac{r_{\rm inner}}{10^{12}\ {\rm cm}}\right)^{3/2}.
\end{eqnarray}
Adopting the masses of the progenitor's He core and the CO to be $5\ {\rm{M_\odot}}$ and $1\ {\rm{M_\odot}}$, respectively and the radius at the base of the envelope to be $10^{12}$ cm, $E_{\rm{orb}}/t_{\rm Kep}$ is on the order of $10^{42}\ {\rm erg\ s^{-1}}$. Thus, for a range of $\alpha\approx 0.1$--$1$ mentioned in \cite{Klencki2021}, 
we find from Figure \ref{fig:tau_max} that this energy injection can indeed result in eruptive mass loss.
However, our results are based on a one-dimensional study assuming spherical symmetry. The above constraint can therefore be somewhat modified since the common envelope process is clearly multi-dimensional.

\subsection{Possible Caveats}
We conclude by brefly discussing several possible caveats of our work.

We have calculated the eruptive mass loss from a RSG using $1$D models under the assumption of spherical symmetry and found conditions for the energy injection to lead to mass eruption. First, these conditions are obtained for a particular RSG model. Though we have extended our results to more general contexts, in reality we should investigate other models to prove whether the conditions really work in general contexts. Nevertheless, we have obtained the values of $E_{\rm{outer}}/t_{\rm{dyn}}$ in equation (\ref{eq:Linj}) for other sample stellar models ($13-20\ M_\odot$) available in CHIPS \citep{Takei21}, and find that they are within $30\%$ of the $15\ M_\odot$ model used in this work. Thus we expect that our conditions will be applicable to RSGs in general.

Second, there is growing evidence for asymmetry in the CSM of some SNe IIn progenitors, e.g., from observations of polarization \citep[e.g.,][]{Burrows1995,Plait1995,Larsson2016}. For example, a disklike CSM was suggested from the optical spectropolarimetry of SN 1998S \citep{Leonard2000}.  {Such strongly aspherical CSM may be difficult to reproduce by mass-loss under the wave heating mechanism, even though multi-dimensional effects such as Rayleigh-Taylor instabilities operate upon energy deposition \citep{Leung2020}. This may hint a different energy injection mechanism at play for the CSM of SN 1998S, such as binary interactions as discussed above. Investigating the outcome of the CSM structure with a more model-agnostic aspherical energy injection would be an interesting future work.}

Third, the energy injection was assumed to occur only once. However past observations of some SNe have found signs of multiple mass eruptions before the terminal explosion exemplified by SN 2009ip \citep[e.g.,][]{Pastorello13,Prieto2013,ofek2013outburst,Margutti2014}. Energy injection can also significantly alter the density structure of the star on the thermal timescale, which might not be recovered at the onset of the next eruption. After the envelope inflates, eruption can become substantially easier due to the weakened gravitational binding of the envelope \citep{Ouchi2019,KS21}. Therefore, if the energy injection is repeated, the condition for eruption can be different for each injection. Such multiple eruptions should be the subject of future research.
\software{ {MESA \citep{Paxton11,Paxton13,Paxton15,Paxton18,Paxton19}, CHIPS \citep{Takei21}, NumPy \citep{Numpy} version 1.22, Matplotlib \citep{matplotlib} version 3.5.}}

\begin{acknowledgments}
The authors thank the anonymous referee for many important comments that improved this manuscript to a great extent. D.T. is supported by the Advanced Leading Graduate Course for Photon Science (ALPS) at the University of Tokyo, and by the JSPS Overseas Challenge Program for Young Researchers. Y.T. is supported by the RIKEN Junior Research Associate
Program. This work is also supported by JSPS KAKENHI grant Nos. JP19J21578, JP20H05639, JP21J13957, MEXT, Japan.
\end{acknowledgments}

\bibliography{references}{}
\bibliographystyle{aasjournal}

\end{document}